\documentstyle[epsfig,12pt]{article}                              
\def\word#1{\,\,\mbox{#1}\,\,}
\def\reff#1{(\ref{#1})}

\def\beq{\begin{equation}}
\def\eeq#1{\label{#1}\end{equation}}

\def\dfrac#1#2{{\displaystyle\frac{#1}{#2}}}

\begin{document}				
\begin{center}
{\bf Classical tests for Weyl gravity: deflection of light and time delay}
\end{center}
\vspace{1em}
\centerline {A. Edery and M. B. Paranjape}
\vspace{1em}
\begin{center}
{\it{Groupe de Physique des Particules, D\'epartement de Physique, 
 \\Universit\'e de Montr\'eal,
C.P. 6128,\\ succ. centreville, Montr\'eal, Qu\'ebec, Canada, H3C 3J7}}
\end{center}

\vspace{3em}
\centerline{\bf Abstract}
\vspace{1em}

{\small Weyl gravity has been advanced in the recent past as an alternative to
General Relativity (GR). The theory has had some success in fitting galactic 
rotation
curves without the need for copious amounts of dark matter. To check the
viability of Weyl gravity, we propose two additional classical tests of the 
theory:
the deflection of light and time delay in the exterior of a static
spherically symmetric source. The result for the deflection of light is 
remarkably
simple: besides the usual positive (attractive) Einstein deflection of
$4GM/r_{0}$ we obtain an extra deflection term of $-\gamma r_{0}$ where
$\gamma$ is a constant and $r_0$ is the radius of closest approach.  With a
negative $\gamma$, the extra term can increase the deflection on large
distance scales (galactic or greater) and therefore imitate the effect of dark
matter. Notably, the negative sign required for $\gamma$ is opposite to the 
sign of $\gamma$ used to fit
galactic rotation curves. The experimental constraints show explicitly that the
magnitude of $\gamma$ is of the order of the inverse Hubble length something
already noted as an interesting numerical coincidence in the fitting of
galactic rotation curves \cite{Mannheim}.

\vspace{3em}
\centerline{\bf I.\quad Introduction}
\vspace{1em}

The higher-derivative conformally invariant Weyl action, the integral of the
square of the Weyl tensor, has attracted much interest as a candidate action for
quantum gravity . Unlike GR, the lack of scale in the theory probably implies that it is
pertubatively renormalizable \cite{Stelle,Tseytlin}. The theory is also 
asymptotically free \cite{Fradkin,Tonin}.   

Weyl gravity, as a classical theory, has attracted less attention because GR has
been so remarkably successful at large distances i.e. on solar system scales, and
therefore there seems no pressing need to study a higher-derivative alternative
classical theory.  However, GR may not be free of difficulties either theoretical or
experimental.  At present, it is faced with one long-standing problem: the
notorious cosmological constant problem \cite{Weinberg} whose solution is not
yet in sight.  There may however be an experimental problem with GR: the
so-called dark matter problem. The clearest evidence for the existence of large
amounts of dark matter comes from the flat rotation curves of galaxies, velocities
of galaxies in clusters and the deflection of light from galaxies and clusters
\cite{Trimble} (for short, we will call these observations ``galactic
phenomenology").  From this evidence, there is a consensus in the astrophysical
community that most of the mass of galaxies (and of our universe) consists of
non-luminous matter.  However, the nature of this dark matter is still unknown
and is one of the great unsolved problems in astrophysics.  At first it was
thought that it may be faint stars or other forms of baryonic matter i.e. the
so-called massice compact halo objects (MACHOS). However, it is safe to say
that observations have obtained much fewer events than required for an
explanation of the galactic phenomenology with a dark halo dominated by
MACHOS \cite{Alcock} (though there is still the possibility that future
experiments might show otherwise). One is then left to consider non-baryonic
forms of dark matter such as massive neutrinos, axions and WIMPS i.e. the
weakly interacting massive particles as predicted for example by supersymmetric
theories. The direct experimental observation of such non-baryonic candidates is
of date singularly lacking ( though many experiments are currently under
development) \cite{Spergel}.  Hence, to date, the nature of the dark matter that is
thought to comprise most of the mass of our universe is still elusive. Is it
possible that the copious amounts of dark matter we are searching for is simply
not there? We believe it is reasonable at this juncture to consider such a 
possibility.

As far as we know, the deviation of galactic rotation curves from the Newtonian
expectation occurs at distances way beyond the solar-system scale
\cite{Sanders}.  In other words, it is a galactic scale phenomena. Newton's gravity
theory, which GR recovers in the non-relativistic weak gravity limit, was
originally formulated to explain solar-system phenomenology and it may be
incorrect to extrapolate this theory to galactic scales.  It has therefore been
suggested by a handful of authors \cite{Milgrom,Mannheim,Sanders} that there
may not be large amounts of dark matter after all and that the ``galactic
phenomenology" may be signaling a breakdown of Newtonian gravity  (and
hence GR) on galactic scales.  

Some authors have therefore proposed alternative classical theories of gravity. 
Most notably there is Milgrom's MOND program \cite{Milgrom}, Mannheim and
Kazanas' Weyl (conformal) gravity program \cite{Mannheim} and Bekenstein and
Sander's scalar-tensor gravity theory \cite{Bekenstein}. In MOND, Newtonian
dynamics are modified at low accelerations typical of orbits on galactic scales. It
has had success in fitting galactic rotation curves without the need for dark
matter \cite{Milgrom,Sanders}. MOND, however, is a non-relativistic theory and
therefore cannot make any predictions on relativistic phenomena such as the
deflection of light, cosmology, etc. In the scalar-tensor theory, it has
been shown that the bending of light cannot exceed that which is predicted by
GR \cite{Bekenstein}, in conflict with the observations i.e. the observed bending
is actually even greater than that predicted by GR. On aesthetic grounds,
conformal gravity is more appealing than other alternative theories because it is
based on a local invariance principle i.e. conformal invariance of the metric. Weyl
gravity encompasses the largest symmetry group which keep the light cones
invariant i.e. the 15 parameter conformal group. It has already been stressed in
the past that unlike Weyl gravity and gauge theories, GR is not based on an
invariance principle. The Principle of General Covariance, which follows from the
Principle of Equivalence, is not an invariance principle. It describes how physical
systems behave in a given arbitrary gravitational field but it does not tell us
much about the gravitational field itself beyond restricting the gravitational
action to a scalar.  The lack of an invariance principle is partly the reason why
guesswork is inevitable in the derivation of Einstein's gravitational field
equations (see \cite{Wein} for details). In contrast, the Weyl action is unique due
to its conformal invariance. Besides its aesthetic appeal, Weyl gravity has many
other attractive features not the least being that it is renormalizable owing to its
lack of length scale. Since the early days of GR, it has been known that the
vacuum GR equations $R_{\mu\nu}=0$ are also vacuum solutions of the Weyl
theory. One therefore expects the Schwarzschild metric to be one possible
solution to the spherically symmetric Weyl vacuum equations. More recently,
Weyl gravity has attracted some interest because it has had reasonable success in
fitting galactic rotation curves without recourse to any dark matter \cite{Mann}.  

The principal reason that Weyl gravity has not received general acceptance is
because some solutions of the classical theory are expected to have no lower
energy bound and therefore exhibit instabilities \cite{Boulware} i.e. runaway
solutions common to higher-derivative theories. For example, there may exist
some Weyl vacuum solutions other than $R_{\mu\nu}=0$ which are not desirable.
Though it has been shown that the Einstein-Hilbert action plus higher-derivative
terms has a well posed initial value problem \cite{Noakes} this has yet to be
shown for the pure fourth order Weyl gravity.  Fortunately, however, the static
spherically symmetric vacuum solutions \cite{Mannheim}, the analog to the
Schwarzschild metric, has been found to be stable and to make important
corrections to the Schwarzschild metric at large distances i.e. it  contains a linear
potential that plays a non-trivial role on galactic scales. It therefore becomes
compelling and interesting to compare Weyl gravity to GR in their classical
predictions.

\vspace{3em}
\centerline{\bf II.\quad Geodesic Equations}
\vspace{1em}
 
Weyl gravity is a theory that is invariant under the conformal transformation
$g_{\mu\nu}(x)\to \Omega^{2}(x)g_{\mu\nu}(x)$ where $\Omega^{2}(x)$ is a finite,
non-vanishing, continuous real function. The metric exterior to a static
spherically symmetric source (i.e. the analog of the Schwarzschild solution in
GR) has already been obtained in Weyl gravity by Mannheim and Kazanas
\cite{Mannheim}. For a metric in the standard form 
\beq 
d\tau^{2}=B(r)\,dt^2 -
A(r)\,dr^2 -r^{2}\left(d\theta^2 + \sin^{2}\theta\,d\varphi^{2}\right) \eeq{ds2} they
obtain the vacuum solutions \beq B(r)=A^{-1}(r)= 1-\dfrac{2\beta}{r} +\gamma\,r
-kr^2 
\eeq{solution} 
where $\beta$,$\gamma$ and $k$  are constants. The authors note that with
$\beta=GM$, the Schwarzschild metric can be recovered on a certain distance
scale (say the  solar system) provided $\gamma$ and $k$ are small enough. The
linear $\gamma$ term would then be significant only on larger distance scales
(say galactic or greater) and hence would deviate from Schwarzschild only on
those scales.  The constant $k$, which should be taken negative, can then be
made even smaller so that the $k r^2$ term becomes significant only on
cosmological scales (in fact, it has been shown \cite{Mannheim} that $k$ is
proportional to the cosmological scalar curvature).  It should be noted that the
solution \reff{solution} is not unique. The Weyl gravitational field equations are
conformally invariant so that any metric which is related to the standard metric
\reff{ds2} by a conformal factor $\Omega^{2}(r)$ is also a valid solution.  This is in
contrast to GR where the Schwarzschild solution is the unique vacuum solution
for a spherically symmetric source. Two metrics that differ by a conformal factor
of course have different curvatures. Remarkably, however, the geodesic
equations for light are conformally invariant. Massive particles, on the other
hand, have geodesics that depend on the conformal factor (though it is 
conceivable to envisage some spontaneous conformal symmetry breaking mechanism which
gives rise to conformally covariant massive geodesics. e.g. see \cite{Mann2}. We
do not entertain conformal symmetry breaking in this paper).  

The geodesic equations along the equatorial plane ($\theta=\pi/2$) for a metric of
the form 
\reff{ds2} are \cite{Wein}
\begin{eqnarray}
r^{2}\dfrac{d\varphi}{dt}&=&J\,B(r)\label{geodesic1}\\
\dfrac{A(r)}{B^{2}(r)}\left(\dfrac{dr}{dt}\right)^{2} +\dfrac{J^2}{r^2}-
\dfrac{1}{B(r)}&=&-E \label{geodesic2}\\ d\tau^{2}&=&E\,B^{2}(r)\,dt^2
\label{geodesic}
\end{eqnarray}
where $E$ and $J$ are constants with $E=0$ for null geodesics (photons) and
$E>0$ for massive particles. The above geodesic equations are only conformally
invariant for photons and therefore two classical tests can be carried out
unambiguously: the deflection of light and the time delay of radar echos.  

\vspace{3em}
\centerline{\bf III.\quad Deflection of Light}
\vspace{1em}

The geodesic equations \reff{geodesic1}-\reff{geodesic} 
enable one to express the angle $\varphi$ as a    
function of $r$ 
\beq
\varphi(r)=\int\dfrac{A^{1/2}(r)}{r^{2}\left(\dfrac{1}{J^{2}\,B(r)}
-\dfrac{E}{J^{2}}-\dfrac{1}{r^2}\right)^{1/2}}\,\,dr.
\eeq{phi}
where the functions $A(r)$ and $B(r)$ are given by \reff{solution}. To do a 
scattering experiment, the light is taken to approach the source 
from infinity. Unlike the Schwarzschild solution where the metric is 
Minkowskian at large distances from the source i.e. $B(r)$ and $A(r)\to 1$ as 
$r\to \infty$, $B(r)$ given by the solution \reff{solution} diverges as 
$r\to \infty$ and we do not recover Minkowski space at large distances. 
However, this is not a problem. At large $r$ it has been shown that the metric 
is conformal to a Robertson Walker metric with three space curvature $K=
-k-\gamma^{2}/4$ \cite{Mannheim}. Hence, at large $r$ the photon is simply 
moving in a
``straight" line in this background geometry (i.e. with $B(r)$ given by
\reff{solution} and $\varphi(r)$ given by \reff{phi}, it is easy to see that
$d\varphi/dr\to 0$ as $r\to\infty$). The photon then deviates from this
``straight" line path as it approaches the source. 

We now substitute the appropriate quantities in Eq.\reff{phi}.  For the photon we
set $E=0$. At the point of closest approach $r= r_{0}$, we have that $dr/d\varphi =0$
and using equations \reff{geodesic} one obtains $(1/J^{2})= B(r_{0})/r_{0}^2 $.
From the solutions \reff{solution} we know that $A^{1/2}(r) = B^{-1/2}(r)$.  The
deflection of the photon as it moves from infinity to $r_{0}$ and off to 
infinity can 
be expressed as 
\beq \Delta\varphi= 2\int_{r_{0}}^{\infty}\left(\dfrac{B(r_{0})}{r_{0}^2}
-\dfrac{B(r)} {r^2}\right)^{-1/2} \,\,\dfrac{dr}{r^2}\,\,\, -\pi 
\eeq{phi3} 
where $\pi$ is
the change in the angle $\varphi$ for straight line motion and is therefore subtracted
out. We now calculate the integral in \reff{phi3} using $B(r)= 1-\dfrac{2\beta}{r}
+\gamma\,r -kr^2$. This yields 
\beq
\int_{r_{0}}^{\infty}\left(\left(1-\dfrac{2\beta}{r_{0}}+\gamma r_{0}\right)
\dfrac{r^{4}}{r_{0}^2}-\gamma r^{3}-r^{2}+2\beta\, r \right)^{-1/2} dr 
\eeq{ellip}
The above integral, being the inverse of the square root of a 
fourth-degree polynomial,
can be expressed in terms of elliptic integrals. However, this is not very 
illuminating. It will prove more instructive to evaluate the integral 
after expanding the integrand 
in some small parameters. 
Note that the constant $k$, important on cosmological scales,  
has cancelled out and does not appear in the integral \reff{ellip}. The
deflection of light is insensitive to the cosmology of the theory and in 
general would not be affected
by a spherically symmetric Hubble flow. On the other hand, the motion of 
massive particles on galactic or greater scales is   
affected by the Hubble flow \cite{Mann,Carmeli}. Hence, the bending of light 
is highly appropriate for testing Weyl gravity.  

We now evaluate the integral \reff{ellip}. It can be rewritten in the form
\beq
\int_{r_{0}}^{\infty}\left(\dfrac{1}{r_{0}^2}-\dfrac{1}{r^2}\right)^{-1/2}
\left\{1-2\beta\left(\dfrac{1}{r_{0}}+\dfrac{1}{r}-\dfrac{1}{r+ r_{0}}\right)+
\dfrac{\gamma r_{0}}{1+ r_{0}/r}\right\}^{-1/2}\,
\dfrac{dr}{r^2}.
\eeq{integ}
After making the substitution
$\sin\theta =r_{0}/r$ the integral becomes
\beq
\int_{0}^{\pi/2} \left[ 1 -\dfrac{2\beta}{r_{0}}\left(1+\sin\theta-        
\dfrac{\sin\theta}{1+\sin\theta}
\right)+\dfrac{\gamma r_{0}}{(1+\sin\theta)}\right]^{-1/2} d\theta
\eeq{calc}
For any realistic situation, such as the bending of light from the sun, galaxies or 
cluster of galaxies the deflection is of the order of arc seconds and therefore the 
parameters
$\beta/r_{0}$ and $\gamma r_{0}$, which measure the deviation from straight line 
motion in Eq. \reff{calc}, must be much less than one. We will therefore expand 
the integrand to first order in the small
parameters $\beta/r_{0}$ and $\gamma r_{0}$. One obtains   
\beq
\int_{0}^{\pi/2} \left[ 1 +\dfrac{\beta}{r_{0}}\left(1+\sin\theta-        
\dfrac{\sin\theta}{1+\sin\theta}
\right)-\dfrac{\gamma r_{0}}{2(1+\sin\theta)}\right] d\theta=
\dfrac{\pi}{2}+\dfrac{2\beta}{r_{0}}-\dfrac{\gamma\,r_{0}}{2}
\eeq{calc2}
The deflection, given by \reff{phi3}, is therefore
\beq
\Delta \varphi = \dfrac{4\beta}{r_{0}} -\gamma\,r_{0}
\eeq{final}
a simple modification of the standard ``Einstein" result of $4GM/r_{0}$ ( where
$\beta=GM$). The constant $\gamma$ must be small enough such that the extra
term $-\gamma\,r_{0}$ is negligible compared to $4GM/r_{0}$ on solar distance
scales. The linear $\gamma$ term, however, can begin to make important
contributions on larger distance scales where discrepencies between experiment
and theory presently exist i.e. the ``Einstein" deflection due to the luminous
matter in galaxies or clusters of galaxies is less than the observed deflection.  Of
course, these discrepencies are usually taken as evidence for the existence of
large amounts of dark matter in the halos of galaxies.  If the extra term
$-\gamma\,r_{0}$ is to ever replace or imitate this dark matter on large distance
scales it would have to be positive (i.e. attractive),  implying that $\gamma$ must
be negative. The sign of $\gamma$ used to fit galactic rotation curves \cite{Mann}
however, is positive ( the reason why the sign of $\gamma$ is different for null and non-relativistic massive geodesics is discussed in the next section on potentials). Therefore there is a glaring incomaptibility between these
two analyses. This means that Weyl gravity does not seem to solve the dark
matter problem, although this does not signal any inconsistency of Weyl gravity
itself.   In addition, the mechanism of conformal symmetry breaking is not well
understood and it must be addressed in more detail before considering massive
geodesics or just mass in general.  The analysis of the deflection of light is more
reliable since it is completely independent of any such conformal symmetry
breaking mechanism. 

\vspace{3em}
\centerline{\bf IV.\ The Potential in Weyl Gravity}
\vspace{1em}

In General Relativity, the Schwarzschild geodesic equations can be viewed as
``Newtonian" equations of motion with a potential (see \cite{Wald}).
In Weyl gravity, a potential can also be extracted from the vacuum
equations and for this purpose it is convenient to define a new ``time"
coordinate $p$ such that $dp=B(r)dt$. The vacuum equations
\reff{geodesic1}-\reff{geodesic} in these new coordinates are
\begin{eqnarray}
r^{2}\dfrac{d\varphi}{dp}&=&J\label{geod1}\\
\dfrac{1}{2}\left(\dfrac{dr}{dp}\right)^{2} +\dfrac{J^2}{2r^2}B(r)-
\dfrac{1}{2}&=&\dfrac{-E\,B(r)}{2}\label{geod2}\\ d\tau^{2}&=&E\,dp^2 .
\label{geod}
\end{eqnarray}
Let $B(r)\equiv 1+2\phi(r)$ where $\phi$ is not necessarily a weak field.  
Equation \reff{geod2} becomes
\beq
\dfrac{1}{2}\left(\dfrac{dr}{dp}\right)^{2} +\dfrac{J^2}{2r^2}+ 
\phi\left(\dfrac{J^2}{r^2} + E\right)=\dfrac{1-E}{2} .
\eeq{geod2x}
The above geodesic equation together with eq.\reff{geod1} can be viewed as a 
particle having energy per unit mass $(1-E)/2$ and angular momentum 
$J$ moving in ordinary mechanics with a potential
\beq
V(r)= \phi\left(\dfrac{J^2}{r^2} + E\right).
\eeq{veff}
The derivative of the potential is
\beq
V'(r) =\dfrac{\beta}{r^2}\left(\dfrac{3J^2}{r^2}+ E\right) + \dfrac{\gamma}{2}
\left(E-\dfrac{J^2}{r^2}\right) -krE.
\eeq{vprime}
where $\phi(r)=-\beta/r + \gamma r/2 -k r^{2}/2$ was used.
There are three terms in Eq. \reff{vprime}: a $\beta$, $\gamma$ and $k$ term respectively. The $k$ term vanishes for null geodesics in agreement with our results on the deflection of light. For massive geodesics the $k$ term is non-zero but is negligible unless one is considering cosmological scales. Hence, this term will be ignored. The factor $3J^{2}/r^{2}+E$ in front of the $\beta$ term
is always positive 
since $E \ge 0$. Therefore, the $\beta$ term is attractive for both massive and
null geodesics (which is the case in GR). On the other hand, 
the factor 
$E-J^{2}/r^{2}$ in front of the $\gamma$ term, can be positive or
negative depending on the physical situation. For a 
non-relativistic particle moving in a weak field, which is the case of 
galactic rotation curves, we obtain $E\approx 1$, $J^{2}/r^{2}\ll 1$, and 
therefore the factor $E-J^{2}/r^{2}$ is positive. For light, E is zero and the 
factor is negative. The potential \reff{veff} is different for non-relativistic particles and light: the $\gamma r$ term in $\phi$ 
contributes 
a linear potential for non-relativistic particles but an inverse $r$ potential
for light.
Their corresponding
derivatives therefore have opposite sign and this explains why  
$\gamma$ obtained through galactic rotation curves has the opposite sign to 
that obtained in the deflection of light. 

Of course, a negative 
$\gamma$ term is not reserved to null geodesics only. Any massive particle 
which is
sufficiently relativistic will also have this 
property. For example consider a particle moving in a weak field $\phi$ with a 
negligible 
``radial velocity" $dr/dp$. One obtains from eq. \reff{geod2x} that 
$J^{2}/r^{2} \approx 1-E-2\phi$ and therefore 
$E-J^{2}/r^{2}\approx 2E+2\phi -1$.
It follows that if a particle is sufficiently relativistic such that $E<1/2 -\phi 
\approx 1/2$ then we obtain a negative $\gamma$ term.

We can actually reproduce the deflection of light result Eq.\reff{final} in a most straightforward way using the potential 
Eq.\reff{veff}. For null geodesics($E=0$) the potential is given by
\beq
V_{null}(r)=\dfrac{-\beta\,J^{2}}{r^{3}}+\dfrac{\gamma\,J^{2}}{2\,r}+\dfrac{-k\,J^{2}}{2}.
\eeq{vnull} 
The deflection by a potential $V(r)$ is obtained by integrating along the 
straight line path the gradient of $V(r)$(in the $\perp$ direction i.e. in
the direction of $r_{0}$). As long 
as the deflection is very small, integrating along the straight line path
instead of the curved path gives the same results.  
The deflection is given by
\beq
\Delta \varphi = \int_{-\infty}^{\infty}\nabla_{\perp} V(r)\,dZ .
\eeq{grad}
where $Z$ is the distance along the
straight line path i.e. $r^{2} =
Z^{2}+ r_{0}^{2}$. In the potential $V_{null}$, the $\gamma$ term is an inverse $r$ potential. This is
the reason why its contribution to the deflection of light Eq.\reff{grad}
is finite and comes with a relative negative sign. If $V_{null}$ had contained a linear potential, the 
integral for the
deflection would diverge, implying that no 
scattering states could exist. 
 
Using $J^{2}=r_{0}^{2}/B(r_{0})$ given in section III and  
$V_{null}$ as the potential, the deflection Eq. \reff{grad} 
yields
\beq
\Delta \varphi= \dfrac{4\beta}{r_{0}}-\gamma\,r_{0}
\eeq{result2}
where only first order terms in $\beta/r_{0}$ and $\gamma\, r_{0}$ were kept. The deflection of light result Eq.\reff{result} is therefore reproduced in a straightforward fashion that allows one to trace clearly the origin of the negative sign in $-\gamma r_{0}$.

\vspace{3em}
\centerline{\bf V.\quad The Weyl Radius}
\vspace{1em}

The geometry of a typical lens system is shown in Fig. 1, below.  
A light ray from a source S is deflected by an angle $\alpha$ at the lens and reaches an observer at O. The angle between the optic axis and the true position of the source is $\beta$ and the angle between the optic axis and the image I is $\theta$. The angular diameter distances between observer and lens, lens and source, and observer and source are $d_{ol},d_{ls}$ and $d_{os}$ respectively. For a spherically symmetric lens, image formation is governed by the one dimensional lens equation 
\beq
\beta=\theta-\alpha(d_{ls}/d_{os}).
\eeq{lens}
 A source is imaged as a ring if the source , the lens and the observer lie on 
a ``straight" line(i.e. $\beta =0$). For an Einstein deflection angle of 
$\alpha=4GM/r_{0}$, the radius of the ring is called the Einstein radius 
and is given by
\beq
\theta_{E}=\left(\dfrac{4GM}{D}\right)^{1/2}
\eeq{radiusE}
where $D\equiv \dfrac{d_{ol}\,d_{os}}{d_{ls}}$ and $M$ is the mass of the lens 
enclosed in the Einstein radius. For a
Weyl deflection angle given by Eq.\reff{final}, the radius of the ring, which we will call
the ``Weyl" radius, can be readily
calculated and yields
\beq
\theta_{w}=\left(\dfrac{4GM}{D+\gamma\, (d_{ol})^{2}}\right)^{1/2} .
\eeq{radiusW}
The above result for the Weyl radius will be used later to obtain an estimate 
for the constant $\gamma$. If the source, lens and observer are not alligned 
in a ``straight" line(i.e. $\beta\ne 0$) then instead of a ring one obtains 
two images, one inside and one outside the Weyl ring. Using the Weyl deflection
angle Eq.\reff{final} and the definitions for the Einstein and Weyl radius, 
the lens equation \reff{lens} gives  
\beq
\beta=\left(1+ n_{\gamma}\right)\theta-\dfrac{4GM}{D\,\theta}
\eeq{image2}
where $n_{\gamma}\equiv \gamma d_{ol}^{2}/D$. The two solutions to the above equation are
\beq
\theta_{\pm}=\dfrac{1}{2(1+n_{\gamma})}\left(\beta\pm \sqrt{\beta^{2}+ 4\theta_{w}^{2}(1+n_{\gamma})^{2}}\right).
\eeq{image}
\begin{figure}
\vspace{-2.cm}
\leavevmode
\epsfxsize=0pt\epsfbox{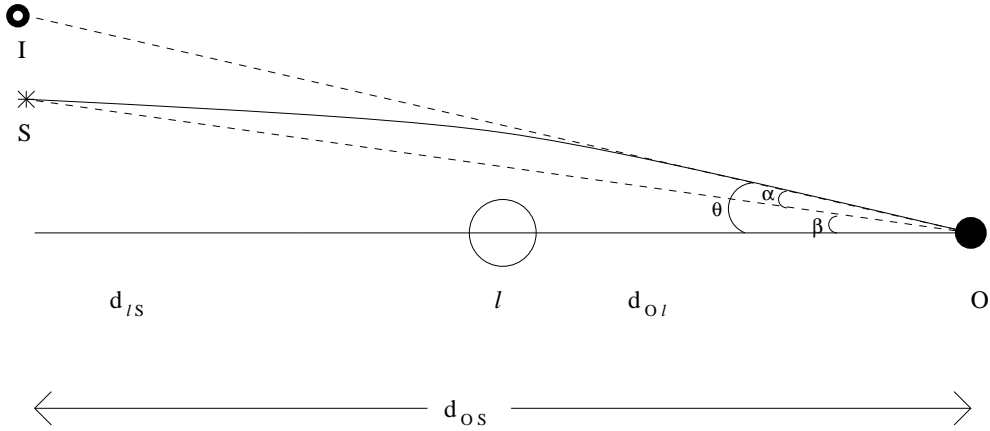}
\vspace{-2.cm}
\caption{Light from the source S bends at the lens {\it l} and arrives at the
observor O who then sees the image I.}
\end{figure}

\vspace{3em}
\centerline{\bf VI.\quad Circular Orbits in Equilibrium}
\vspace{1em}

In the Schwarzschild metric, it is known that photons do not have circular 
orbits with stable equilibrium but have one unstable equilibrium at the radius 
$r=3GM$. We now determine the radii of equilibrium for photons in the Weyl
vacuum solution \reff{solution}. The geodesic equation of interest is
Eq. \reff{geodesic2} where we substitute $E=0$ for photons and set $dr/dt$ to 
zero at the radius of orbit $r=R$. Equation \reff{geodesic2} becomes
\beq
\dfrac{J^{2}}{R^{2}} - \dfrac{1}{B(R)} = 0.
\eeq{circle}
For equilibrium, the derivative of the LHS of \reff{circle} at $r=R$ must vanish and 
we obtain
\beq
\dfrac{-2J^2}{R^3} + \dfrac{B^{'}(R)}{B^{2}(R)}=0
\eeq{equi}
With  $J^2$ given by \reff{circle} and $B(r)$ given by \reff{solution}, 
equation \reff{equi} becomes 
\beq
\gamma R^2 + 2R - 6GM =0 
\eeq{eqroots}
where $\beta =GM$ was used. Note that the constant $k$ has again cancelled 
out. The two solutions to equation \reff{eqroots} are
\beq
R \simeq 3GM \word{and} R\simeq -2/\gamma
\eeq{roots}
where it has been assumed that $|\beta\gamma|<<1$. We see that besides the
$R=3GM$ solution a second equilibrium exists at $R=-2/\gamma $ if $\gamma$ is
negative. By differentiating equation \reff{eqroots} we see that this second
equilibrium is a stable one while the first is an unstable one as in the
Schwarzschild case. This stable equilibrium provides us with a natural length
scale i.e. a scale which determines the ``region of influence" of a particular
localized source in contrast to the background or global aspects. A length scale
of this sort is probably necessary if we ever want to develop a concept of
``energy of an isolated system" in Weyl gravity. In the Scwarzschild case, the
metric tends towards Minkowski space in the limit $r\to\infty$ and a Gauss's law
formulation of total energy of an ``isolated" system is possible. In the Weyl case
we obtain a metric conformal to a Robertson-Walker spacetime in the limit
$r\to\infty$. We therefore need a natural cut-off radius at which the influence of
the specific source in question ceases and the global aspects take over. Indeed,
we have shown that the constant $k$, which is proportional to the cosmological
curvature, plays no role in determining the radius of stable equilibrium and lends
support to the idea that the stable radius is determined by the localized source.
Hence, from the arguments above, a negative $\gamma$ is desirable.
                    
\vspace{3em}
\centerline{\bf VII.\quad Time Delay}
\vspace{1em}

We now calculate the time taken by a photon for a trip between any two 
points in a gravitational field produced by a central mass. We expect modifications to the standard GR result when the radius of closest approach to the central mass is on the order of galactic scales. The equation governing the time evolution of orbits is Eq. \reff{geodesic2}, 
with $E=0$ for light. At the point of closest approach $r=r_{0}$, $dr/dt =0$ so 
that Eq. \reff{geodesic2} gives $J^{2}=r_{0}^{2}/B(r_{0})$. The time for light
to travel from $r_{0}$ to $r_{1}$, given by Eq. \reff{geodesic2}, is 
\beq
t= \int_{r_{0}}^{r_{1}} \left(\dfrac{A(r)/B(r)}{1-\dfrac{B(r)\,r_{0}^{2}}{B(r_{0})\,r^{2}}}
\right)^{1/2} dr
\eeq{time}
We evaluate the above integral with $A(r)$ and $B(r)$ given by 
Eq. \reff{solution}. This yields
\beq                                                                              
t= \int_{r_{0}}^{r_{1}} 
\dfrac{r(1-2\beta/r
+\gamma\,r-k\,r^{2})^{-1}(1-2\beta/r_{0} +\gamma\,r_{0} -k\,r_{0}^{2})^{1/2}}
{\sqrt{r^{2}-r_{0}^{2}}\left[1-
\dfrac{2\beta}{r_{0}}\left(1+\dfrac{r_{0}^{2}}
{r(r+r_{0})}\right) +\gamma
r_{0}\left(\dfrac{r}{r+r_{0}}\right)\right]^{1/2}}\,dr
\eeq{intime}
We can expand the above integral to first order in the parameters 
$\beta/r$, $\gamma\, r$ and $k\,r^{2}$ which are much less than 1 within the
usual limits of integration. 
To first order in the parameters, the integral \reff{intime} yields
\beq
t\simeq \int_{r_{0}}^{r_{1}} r\left[
(1-\dfrac{1}{2}kr_{0}^{2}) +\dfrac{2\beta}{r} + \dfrac{\beta\,r_{0}}{r(r+r_{0})}
 -\gamma\,r +\dfrac{\gamma\,r_{0}^{2}}{2(r+r_{0})} +
k\,r^{2}\right]\dfrac{dr}{\sqrt{r^{2}-r_{0}^{2}}}
\eeq{int2}
There are six elementary integrals to evaluate above. The result is
\begin{eqnarray}
t&\simeq&\sqrt{r_{1}^{2}-r_{0}^{2}} +
2\beta\ln\left(\dfrac{r_{1}+\sqrt{r_{1}^{2}-r_{0}^{2}}}{r_{0}}\right)
+
\beta\,\sqrt{\dfrac{r_{1}-r_{0}}{r_{1}
+r_{0}}} \nonumber\\
&-&\dfrac{\gamma}{2}\left(\dfrac{r_{1}^{3}-r_{0}^{3}}
{\sqrt{r_{1}^{2}-r_{0}^{2}}}\right) 
+ \dfrac{k}{6}(2r_{1}^{2}+r_{0}^{2})\sqrt{r_{1}^{2}-r_{0}^{2}}.
\label{result}
\end{eqnarray} 
The leading term is identified as the time for light to travel in a straight 
line in Minkowski space (where $\beta=\gamma=k=0$) and we recognize the
$\beta$ terms as the standard ``Shapiro" time delay. The $\gamma$ and $k$ 
terms evidently produce a modification of the time delay. We see that the 
effect of the $\gamma$ term is to increase the time delay if $\gamma$ is 
negative and to decrease it if $\gamma$ is positive. 

\vspace{2em}
\centerline{\bf VIII.\quad Constraints on $\gamma$ from Experiments}
\vspace{1em}

\noindent{\it A.\quad Solar Gravitational Deflection}
\vspace{1em}

In solar experiments, the sun can be treated as a point mass and no 
lens model is required. To date, the best measurements
on the deflection of light from the sun were obtained using 
radio-interferometric methods and verified Einstein's prediction to within 1
\%.
The measured deflection at the solar limb was $1.761\pm 0.016$ arc sec
\cite{Sramek} compared to Einstein's prediction of $4GM_{\odot}/R_{\odot}=1.75$
arc sec. 
Using the Weyl deflection angle Eq.\reff{final} these  measurements constrain
the constant $\gamma$ to the range 3.45x $10^{-19} cm^{-1}\ge \gamma \ge$ -1.87x
$10^{-18} cm^{-1}$. 
Clearly, the solar gravitational deflection experiments 
constrain strongly the order of magnitude of $\gamma$ but leave open the 
possibility for a positive or negative $\gamma$.

\vspace{3em}
\noindent{\it B.\quad Signal retardation by solar gravity}
\vspace{1em}    
  
The results of the Viking Relativity Experiment published in 
1979 \cite{Shapiro} confirmed the ``Shapiro" time delay on solar system scales 
to an accuracy of 0.1\%. For example, a ray that leaves the 
earth, grazes the sun, reaches Mars and comes back would have a time delay of 
$248 \pm 0.25 \mu s$ where the $248 \mu s$ is the exact prediction of the 
``Shapiro" time delay and the uncertainty $\pm 0.25 \mu s$ can be used to
constrain $\gamma$. At superior conjunction, the radius of the sun to the Earth
, $r_{e}$, and to Mars, $r_{m}$, are much greater than the radius of the sun 
$R_{\odot}$ so that $r_{0}$ can be neglected in the factor in front of $\gamma$
in Eq.\reff{result}. We therefore have $-\gamma (r_{e}^{2}+r_{m}^{2})=\pm
0.25$x$10^{-6} s$. This constrains $\gamma$ to the range $|\gamma |\le 1.02$  
x$10^{-23}cm^{-1}$. This is roughly five orders of magnitude better than the
constraint on $\gamma$ from solar deflection experiments but does not allow
us to draw any conclusions on the sign of $\gamma$. 

\vspace{3em}
\noindent{\it C.\quad Deflection of light by galaxies and clusters}
\vspace{1em}

One should expect measurements on the deflection of light by galaxies 
and clusters to determine the most accurate value for $\gamma$ because it is on
those scales where the $\gamma$ term plays a significant role. However, the 
interpretation of the experimental data on those scales is more difficult than 
in the solar system because galaxies
and clusters have unknown matter distributions and in general cannot be 
assumed to be either point masses or spherically symmetric. A parametrized lens 
model
is therefore required for each case of gravitational lensing. For example, to understand the time delay in the gravitational lens 0957+561,
one has to describe not only the distribution of the lensing galaxy G1 but also the effect of the
surrounding cluster. Nonetheless, it
has been pointed out \cite{Schneider,Narayan} that though a spherically
symmetric lens model is an idealization, it is a good first approximation in 
understanding the large arcs that are observed in clusters and is
extremely useful in obtaining the same order of magnitude results as the more
realistic case. We will therefore use data on the large arcs found in clusters 
to obtain a value for $\gamma$ with confidence that the order of magnitude is 
correct.
To constrain
$\gamma$ beyond ``order of magnitude" accuracy, a detailed 
lens model for each cluster is required and would take us beyond the scope of the 
present paper. 

In a 
spherically symmetric lens, the radius of the tangentially oriented large arcs,
$\theta_{arc}$, is a good estimate of the radius which occurs at $\beta =0$ in 
the lens equation Eq. \reff{lens}. In GR, the radius of the arc is therefore 
interpreted as the Einstein radius where $M$ is the total mass($M_{total}$) 
i.e. the sum of the luminous and presumed dark matter. In the context of Weyl
gravity, the {\it same} arc is to be interpreted as the Weyl radius, with $M$
representing only the luminous matter($M_{L}$) and $\gamma$ a constant to be
determined. Using equations \reff{radiusE} and \reff{radiusW} for the Einstein
and Weyl radius respectively and equating them for the same observed arc, one
obtains 
\beq
\gamma= \left(\dfrac{d_{os}}{d_{ls}\,d_{ol}}\right)\left(\dfrac{M_{L}}{M_{total}}-1\right).
\eeq{gamma1}
In experiments one measures the redshifts $Z_{l}$ and $Z_{s}$ in the spectrum 
of the light reaching us from the lens(i.e. the cluster) and source 
respectively. 
We define a dimensionless quantity(often called the angular size distance)  $y\equiv (1+Z)d/L_{H_{0}}$ 
where $L_{H_{0}}\equiv c/H_{0}$ is the Hubble length and $d$ are the angular diameter distances 
that appear in Eq.\reff{gamma1}(we have reinserted the speed of light c for clarity). The values of $y$ are related to redshifts by \cite{Peebles}
\beq
y_{ox} = \left\{ \begin{array}{r@{\word{for}}l}
\dfrac{Z_{x}(1+Z_{x}/2)}{1+Z_{x}}& \Omega = 0\\
2-\dfrac{2}{\sqrt{1+Z_{x}}}& \Omega = 1\end{array} \right.
\eeq{yox1}
\beq
y_{ls} = \left\{ \begin{array}{r@{\,\,\word{for}}l}
y_{os}(1+y_{ol}^{2})^{1/2}-y_{ol}(1+y_{os}^{2})^{1/2}& \Omega = 0\\
y_{os}-y_{ol}\quad\quad & \Omega = 1\end{array} \right.   
\eeq{yox2}     
where $x$ represents either the lens $l$ or the source $s$ and $\Omega$ is the 
cosmological density parameter(we have taken $\Omega=0$ and
$\Omega=1$ as examples though we will see that $\gamma$ is insensitive to
$\Omega$). Equation \reff{gamma1} can be rewritten in terms of 
$y$ and yields 
\beq
\gamma= \dfrac{1}{L_{H_{0}}}\left(\dfrac{y_{os}}{y_{ls}\,y_{ol}}\right)
\left(\dfrac{M_{L}}{M_{total}}-1\right)
\eeq{gamma2}
To obtain a value for $\gamma$
reliable data on the redshifts $Z_{l}$ and $Z_{s}$ is required as well as
values for the 
mass-to-light ratios of clusters derived from gravitational lensing.
Fortunately, such
data exists for many large arcs in clusters.
Before looking at the data it is important to note that the mass-to-light ratio
is large for a typical
cluster($>$100) and therefore $M_{L}/M_{total}<<1$. It follows that the factor
$(M_{L}/M_{total} - 1)$ will not differ from cluster to cluster. Data is shown 
below (taken from \cite{Narayan2}, \cite{Pello}, \cite{Petrosian}) for 
different clusters with the value of $\gamma$ calculated in each case. 

\vspace{2em}
\begin{tabular}{c|c|c|c|c|c}
Cluster & $Z_{l}$ & $Z_{s}$ & $M_{L}/M_{total}$ & 
$\left.\dfrac{y_{os}}{y_{ls}\,y_{ol}}\right|_{\Omega = 1,0}$ & $\left.\gamma
\right|_{\Omega =1,0}$\\ 
\hline
A370  & 0.375 & 0.724 & $\approx {1}/{200}$ & 6.877, 7.765  &
-6.83, -7.72 \\
A2390  & 0.231 & 0.913 &$\approx {1}/{120}$ &  7.885, 7.308 &  
-7.82, -7.25 \\
Cl2244-02 & 0.331 & 0.83 &$< {1}/{100}$ &7.68, 6.87 &-7.60, -6.80 
\end{tabular}

\vspace{2em}
\noindent where $\gamma$ is in units of the inverse Hubble length, $1/L_{H_{0}}$ 
(which is equal to $(H_{0}/100)\times$\hfil \\ $1.08 \times 10^{-28}cm^{-1}$. As can be seen from the data,
$\gamma$ is negative, reasonably constant from one cluster to the next and its order of magnitude is clearly the inverse Hubble length. Interestingly enough, the value for
$\gamma$ obtained by Mannheim and Kazanas \cite{Mannheim} in the context of galactic rotation curves is of the same order of magnitude. Though no use of 
Hubble's constant was made in their 
calculation, the authors recognized that $\gamma$ was close numerically to the inverse Hubble length.
Lensing data in clusters confirms via Equation\reff{gamma2} that $\gamma$ 
is indeed dependent on Hubble's constant. In conclusion, we obtain the same order of
magnitude for $\gamma$ as in galactic rotation curves but with opposite sign. This discrepency merits further investigation.

\end{document}